\documentclass[twocolumn,secnumarabic,amssymb, nobibnotes, showpacs, aps, prl]{revtex4-1}
\usepackage{graphicx}

\usepackage{epsfig}

\usepackage{lineno}
\begin{document}
%\linenumbers

\title{The $\Lambda\Lambda$ Correlation Function in Au+Au collisions at $\sqrt{s_{NN}}=$ 200 GeV}
\author{
L.~Adamczyk$^{1}$,
J.~K.~Adkins$^{23}$,
G.~Agakishiev$^{21}$,
M.~M.~Aggarwal$^{35}$,
Z.~Ahammed$^{53}$,
I.~Alekseev$^{19}$,
J.~Alford$^{22}$,
C.~D.~Anson$^{32}$,
A.~Aparin$^{21}$,
D.~Arkhipkin$^{4}$
% ...
}

\address{$^{1}$AGH University of Science and Technology, Cracow, Poland}
% ...
\address{$^{4}$Brookhaven National Laboratory, Upton, New York 11973, USA}
% ...
\address{$^{19}$Alikhanov Institute for Theoretical and Experimental Physics, Moscow, Russia}
% ...
\address{$^{21}$Joint Institute for Nuclear Research, Dubna, 141 980, Russia}
\address{$^{22}$Kent State University, Kent, Ohio 44242, USA}
\address{$^{23}$University of Kentucky, Lexington, Kentucky, 40506-0055, USA}
% ...
\address{$^{32}$Ohio State University, Columbus, Ohio 43210, USA}
% ...
\address{$^{35}$Panjab University, Chandigarh 160014, India}
% ...
\address{$^{53}$Variable Energy Cyclotron Centre, Kolkata 700064, India}
% ...
\collaboration{STAR Collaboration}\noaffiliation

\date{December 9, 2014}%

\begin{abstract}
We present $\Lambda\Lambda$ correlation measurements in heavy-ion collisions for Au+Au collisions at $\sqrt{s_{NN}}= 200$ GeV using the STAR experiment at the Relativistic Heavy-Ion Collider (RHIC). The Lednick\'{y}-Lyuboshitz analytical model has been used to fit the data to obtain a source size, a scattering length and an effective range. Implications of the measurement of the $\Lambda\Lambda$ correlation function and interaction parameters for di-hyperon searches are discussed.
\end{abstract}
\maketitle

%\section{Introduction}

Measurements of the correlation function for a pair of particles with small relative momenta have been used to obtain insight into the geometry and lifetime of the particle-emitting source in relativistic heavy-ion collisions~\cite{HBT}. The two-particle correlation function is not only sensitive to the distribution of the separation of emission points, but also to the effects from Quantum Statistics (QS) and to the Final-State Interactions (FSI).  For two-particle systems where the final-state interactions are well known, information about both temporal and spatial separation distributions can be obtained using the two-particle correlation function~\cite{HBT,HBT2}. If one has an idea of the source size, one could use it to determine the FSI between two particles for which the correlation function is measured. In this paper we have used $\Lambda\Lambda$ correlation measurements to determine FSI between $\Lambda\Lambda$ which is not well known experimentally.
 
The $\Lambda\Lambda$ correlation function is also relevant for searching for the $H$-dibaryon, a six-quark state predicted by Jaffe~\cite{Jaffe}.  Recent lattice QCD calculations from the HAL~\cite{HALQCD} and NPLQCD~\cite{NPLQCD} collaborations indicate the possible existence of a bound $H$-dibaryon, where the calculations assumed a pion mass above the physical mass.  The production rate for the hypothesized $H$-dibaryon depends on the collision evolution dynamics as well as on its internal structure.  It is believed that the most probable formation mechanism for the $H$-dibaryon would be through coalescence of $\Lambda\Lambda$ and/or $\Xi N$ at a late stage of the collision process, or through coalescence of six quarks at an earlier stage of the collision~\cite{ExHIC}.  A measurement of the $\Lambda\Lambda$ interaction is important for understanding the equation of state of neutron stars~\cite{hyperonstar}.  Moreover at high densities, an attractive $\Lambda\Lambda$ interaction could lead to formation of $H$-matter or strangelets in the core of moderately dense neutron stars~\cite{strangestar, strangematter}. 

At present, the constraint on the binding energy of the $H$-dibaryon comes from double $\Lambda$ hypernuclei (NAGARA event)~\cite{Nagara}, which allows the possibility of a weakly bound $H$-dibaryon or a resonance state~\cite{akira_hyp2012}. The resonance state is expected to decay into $\Lambda\Lambda$  and would be observed as a bump in the $\Lambda\Lambda$ invariant mass spectrum or observed as a peak-like structure in two-particle correlations~\cite{Muller}.

Dedicated measurements have been performed to look for the $H$-dibaryon signal, but its existence remains an open question~\cite{KEK, E910, kTeV}. The STAR experiment has searched for strangelet production close to the beam rapidity at RHIC and has reported an upper limit for strangelets~\cite{strangelet}. The NA49 experiment at the SPS attempted to measure the $\Lambda\Lambda$ correlation function in heavy-ion collisions, but their statistics were  insufficient to draw physics conclusions~\cite{NA49}. The observed high yield of multi-strange hyperons in central nucleus-nucleus collisions at RHIC~\cite{strangebaryon} and recent high-statistics data for Au+Au collisions at RHIC provide a unique opportunity to study $\Lambda\Lambda$ correlations and search for exotic particles like the $H$-dibaryon. In this Letter, we present the first measurement of the $\Lambda\Lambda$ correlation function in heavy-ion collisions, for Au+Au collisions at $\sqrt{s_{NN}}= 200$ GeV using the STAR experiment at the RHIC.

\begin{figure}[hbt]
\begin{center}

\epsfxsize = 3.4in
\epsfysize = 2.7in
\epsffile{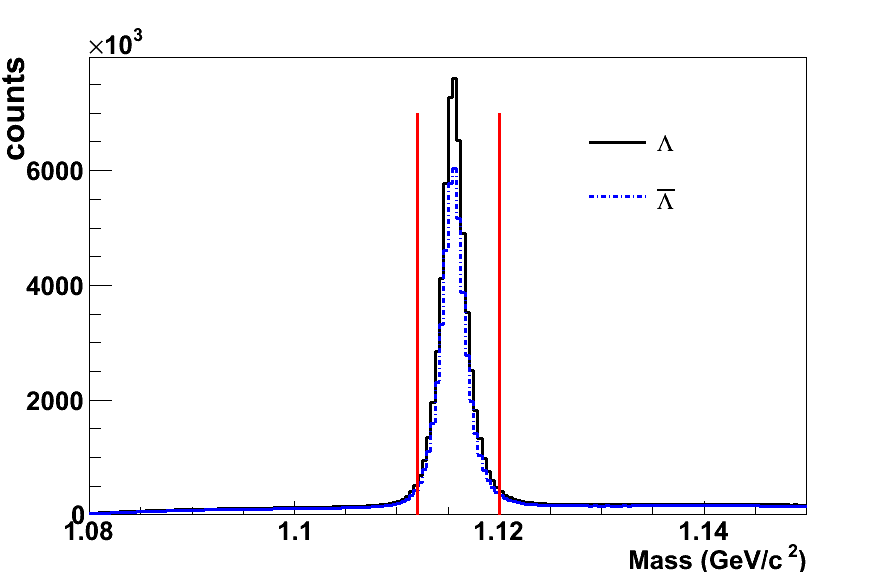} 
\end{center}
\caption{
(color online). The invariant mass distribution for $\Lambda$ and $\bar{\Lambda}$ produced in Au+Au collisions at $\sqrt{s_{NN}} = 200$ GeV, for $0-80$\% centrality. The $\Lambda$ ($\bar{\Lambda}$) candidates lying in the mass range 1.112 to 1.120 GeV$/c^2$, shown by solid red vertical lines, were selected for the correlation measurement.}
\label{fig:IM}
\end{figure}

STAR is a multi-purpose experiment at RHIC with full azimuthal coverage. The Time Projection Chamber~\cite{STAR_det} was used for tracking and particle identification in the pseudorapidity range $|\eta|< 1$. Approximately $2.87 \times 10^8$ events from 2010 and $5.0\times 10^8$ events from 2011 were analyzed. 
To suppress events from collisions with the beam pipe (radius 3.95 cm), the reconstructed primary vertex was required to lie within a 2 cm radial distance from the center of the beam pipe. In addition, the $z$-position of the vertex was required to lie within $\pm 30$ cm of the center of the detector. 
The decay channel $\Lambda \rightarrow p\pi$ with branching ratio $63.9 \pm 0.5$\% was used for reconstruction of the $\Lambda$~\cite{PDG}. The $\Lambda$ ($\bar{\Lambda}$) candidates were formed from pairs of $p$ ($\bar{p}$) and $\pi^{-}$ ($\pi^{+}$) tracks whose trajectories pointed to a common secondary decay vertex which was well separated from the primary vertex.  The decay length (DL) of a $\Lambda$ candidate was required to be more than 5 cm from the primary vertex. The DL cut did not correspond to a hard cut-off in momentum and it was based on the requirement for high purity of the $\Lambda$ sample as well as reasonable efficiency. The distance of closest approach (DCA) to the primary vertex was required to be within 0.4 cm. The invariant mass distribution of the $\Lambda$ ($\bar{\Lambda}$) candidates at $0-80$\% centrality under these conditions as shown in Fig.~\ref{fig:IM} has an excellent signal ($S$) to background ($B$) ratio of $S/(S+B)\sim 0.97$. The solid (dashed) histogram is for $\Lambda$ ($\bar{\Lambda}$) candidates. All candidates with invariant mass between 1.112 and 1.120 GeV$/c^2$ were considered. %In order to increase statistics, results for $\Lambda$ and $\bar{\Lambda}$ were combined. 

The two-particle correlation function is defined as 

\label{myeq}
\begin{equation}
C_{\rm{measured}}(Q) = \frac{A(Q)}{B(Q)},
\end{equation}

\noindent
where $A(Q)$ is the distribution of the invariant relative momentum, $Q=\sqrt{-q^\mu q_\mu}$, where $q^\mu = p^\mu_1 - p^\mu_2$, for a pair of $\Lambda\,(\bar{\Lambda})$ from the same event. $B(Q)$ is the reference distribution generated by mixing particles from different events with approximately the same vertex position along the $z$-direction. The same single-particle cuts were applied to individual $\Lambda$s for the mixed-event pairs. Correlations between a real $\Lambda$ and a false $\Lambda$ candidate reconstructed from a pair that shares one or two daughters with the real $\Lambda$ were avoided by removing any $\Lambda$ pair with a common daughter. Possible two-track biases from reconstruction were studied by evaluating correlation functions with various cuts on the scalar product of the normal vectors to the decay plane of the $\Lambda$s and on the radial distance between $\Lambda$ vertices in a given pair. No significant change in the correlation function has been observed due to these tracking effects. Each mixed event pair was also required to satisfy the same pair-wise cuts applied to the real pairs from the same event. The efficiency and acceptance effects canceled out in the ratio $A(Q)/B(Q)$. Corrections to the raw correlation functions were applied according to the expression

\label{myeq2}
\begin{equation}
C'(Q) = \frac{C_{\rm{measured}}(Q)-1}{P(Q)}+1, 
\end{equation}
\noindent 
where the pair purity, $P(Q)$, was calculated as a product of $S/(S+B)$ for the two $\Lambda$s of the pair. The pair purity is 92\% and is constant over the analyzed range of invariant relative momentum. 

The selected sample of $\Lambda$ candidates also included secondary $\Lambda$s, i.e. decay products of $\Sigma^0,\, \Xi^-$ and $\Xi^0$, which were still correlated because their parents were correlated through quantum statistics and emission sources. Toy model simulations have been performed to estimate the feed-down contribution from $\Sigma^0\Lambda$, $\Sigma^0 \Sigma^0$ and $\Xi^- \Xi^-$. The $\Lambda$, $\Sigma$ and $\Xi$ spectra have been generated using a Boltzmann fit at midrapidity ($T=335$ MeV~\cite{strangebaryon}) and each pair was assigned a weight according to quantum statistics. The pair was allowed to decay into daughter particles and the correlation function was obtained by the mixed-event technique. The estimated feed-down contribution was around 10\% for $\Sigma^0\Lambda$, around 5\% for $\Sigma^{0}\Sigma^{0}$ and around 4\% for $\Xi^{-}\Xi^{-}$. Thermal model studies have shown that only 45\% of the $\Lambda$s in the sample are primary~\cite{therminator}. However, one needs to run afterburners to determine the exact contribution to the correlation function from feed-down, which requires knowledge of final-state interactions. The final-state interaction parameters for $\Sigma^{0}\Sigma^{0}$, $\Sigma^{0}\Lambda$ and $\Xi\Xi$ interactions are not well known, which makes it difficult to estimate feed-down using a thermal model~\cite{therminator}. Therefore, to avoid introducing large systematic uncertainties from the unknown fraction of aforementioned residual correlations, the measurements presented here are not corrected for residual correlations.

\begin{figure}[hbt]
\begin{center}

\epsfxsize = 3.4in
\epsfysize = 2.5in
\epsffile{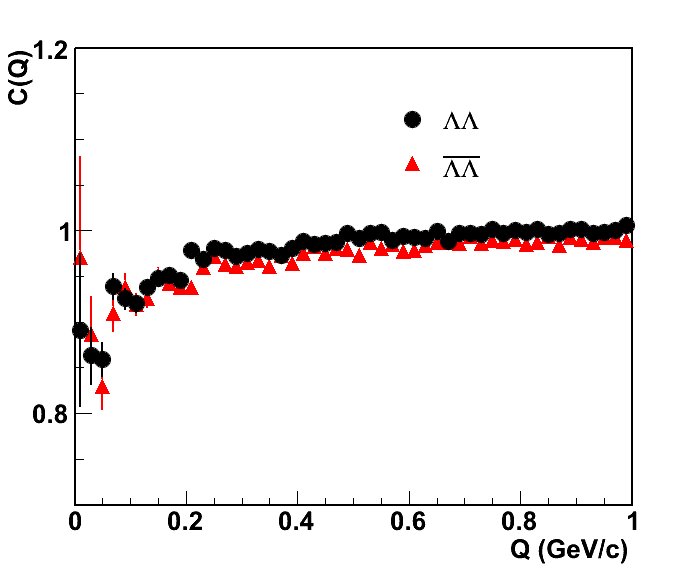} 
\end{center}
\caption{ 
(color online). The $\Lambda\Lambda$ and $\bar{\Lambda}\bar{\Lambda}$ correlation function in Au+Au collisions at $\sqrt{s_{NN}} = 200$ GeV, for 0-80\% centrality. The plotted errors are statistical only.}
\label{fig:CF_lmanlm}
\end{figure}

The effect of momentum resolution on the correlation functions has also been investigated using simulated tracks from $\Lambda$ decays, with known momenta, embedded into real events. Correlation functions have been corrected for momentum resolution using the expression
\begin{equation}
C(Q) = \frac{C'(Q) C_{\rm in}(Q)}{C_{\rm res}(Q)}, 
\end{equation}
\noindent 
where $C(Q)$ represents the corrected correlation function, and $C_{\rm in}(Q) / C_{\rm res}(Q)$ is the correction factor. $C_{\rm in}(Q)$ was calculated without taking into account the effect of momentum resolution and $C_{\rm res}(Q)$ included the effect of momentum resolution applied to each $\Lambda$ candidate.  More details can be found in Ref.~\cite{pionHBT}. The impact of momentum resolution on correlation functions was negligible compared with statistical errors. Figure~\ref{fig:CF_lmanlm} shows the experimental $\Lambda\Lambda$ and $\bar{\Lambda}\bar{\Lambda}$ correlation function after corrections for pair purity and momentum resolution for 0-80\% centrality Au+Au collisions at $\sqrt{s_{NN}} = 200$ GeV. The $\bar{\Lambda}\bar{\Lambda}$ correlation function is slightly lower than the ${\Lambda}{\Lambda}$ correlation function, although within the systematic errors. Noting that the correlations $C(Q)$ in Fig.~\ref{fig:CF_lmanlm} are nearly identical for $\Lambda$ and $\bar{\Lambda}$, we have chosen to combine the results for $\Lambda$ and $\bar{\Lambda}$ in order to increase the statistical significance.

\begin{figure}[h]
\begin{center}

\epsfxsize = 3.3in
\epsfysize = 2.7in
\epsffile{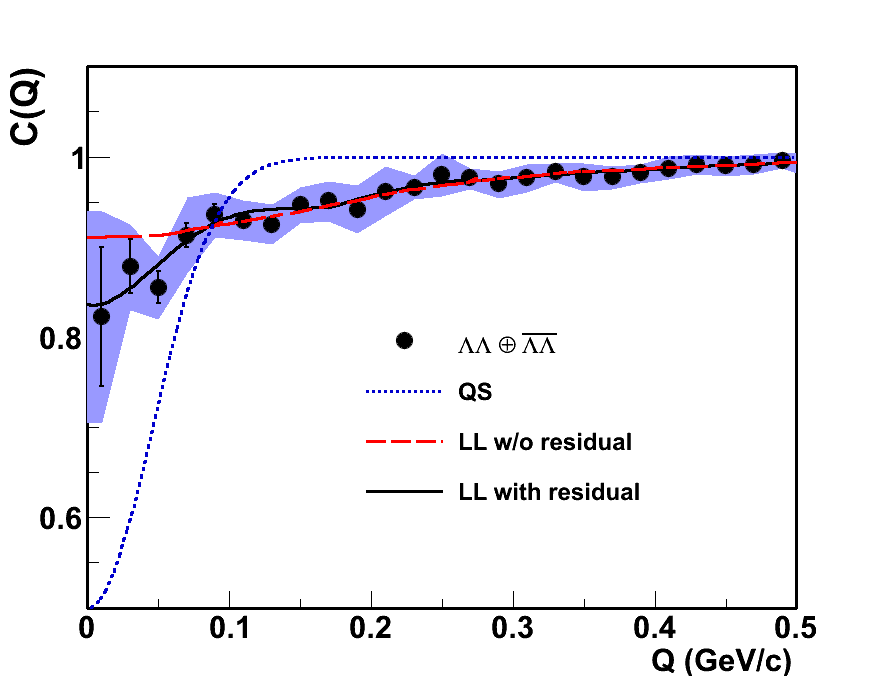} 
\end{center}
\caption{
(color online). The combined $\Lambda\Lambda$ and $\bar{\Lambda}\bar{\Lambda}$ correlation function for 0-80\% centrality Au+Au collisions at $\sqrt{s_{NN}} = 200$ GeV. Curves correspond to fits using the Lednick\'{y}-Lyuboshitz (LL) analytical model with and without a residual correlation term~\cite{Lednicky}. The dotted line corresponds to quantum statistics with a source size of 3.13 fm. The shaded band corresponds to the systematic error.  
}
\label{fig:CF_true}
\end{figure}

 The combined $\Lambda\Lambda$ and $\bar{\Lambda}\bar{\Lambda}$ correlation function for 0-80\% centrality is shown in Fig.~\ref{fig:CF_true}. The systematic errors were estimated by varying the following requirements for the selection of $\Lambda$: DCA, DL and mass range, which affect the signal-to-background ratio. Systematics from cuts on the angular correlation of pairs were also studied that may affect correlations at small relative momentum. The systematic uncertainties from different sources were then added in quadrature. The combined systematic error is shown separately as a shaded band in Fig.~\ref{fig:CF_true}. If there were only antisymmetrization from quantum statistics, a $\Lambda\Lambda$ correlation function of 0.5 would be expected at $Q=0$. The observed pair excess near $C(Q=0)$ compared to 0.5 suggests that the $\Lambda\Lambda$ interaction is attractive, however as mentioned earlier, the data are not corrected for residual correlations and those effects can give rise to this excess. In Fig.~\ref{fig:CF_true}, the dotted line corresponds to quantum statistics.  

The Lednick\'{y} and Lyuboshitz analytical model~\cite{Lednicky} relates the correlation function to source size and also takes into account the effect of the strong final-state interactions (FSI). The following correlation function is used to fit the experimental data 

\begin{eqnarray}
\label{ll_eq}
\nonumber
C(Q)& = &N\large{[}1+\lambda\Large{(}-\frac{1}{2}\exp(-r_{0}^{2}Q^{2})+\\
\nonumber
&& \frac{1}{4}\frac{\vert f(k)\vert^{2}}{r_{0}^{2}} (1- \frac{1}{2\sqrt{\pi}}\frac{d_{0}}{r_{0}})+ \\ 
\nonumber
&& \frac{{\rm Re\,} f(k)}{\sqrt{\pi}r_0}F_1(Qr_0) - \frac{{\rm Im\,} f(k)}{2r_0}F_2(Qr_0)\Large{)}+\\
&& a_{\rm res}\exp(-r_{\rm res}^2 Q^2)\large{]}, 
\end{eqnarray}
\noindent 
where $k=Q/2$, $F_1(z) = \int_0^1 e^{x^2 -z^2}/z\, dx$ and $F_2(z)= (1-e^{-z^2})/z$ in Eq.~(\ref{ll_eq}). The scattering amplitude is given by 

\begin{equation}
f(k)=(\frac{1}{f_{0}}+\frac{1}{2}d_{0}k^{2}-ik)^{-1},
\end{equation}
\noindent 
where $f_{0}=a_{0}$ is the scattering length and $d_{0}=r_{\rm eff}$ is the effective range. Note that a universal sign convention is used rather than the traditional sign convention for the s-wave scattering length, $a_{0} = -f_{0}$ for baryon-baryon systems. More details about the model can be found in Ref.~\cite{Lednicky}. The free parameters of the LL model are normalization ($N$), a suppression parameter ($\lambda$), an emission radius ($r_{0}$), scattering length ($a_{0}$) and effective radius ($r_{\rm eff}$). In the absence of FSI, $\lambda$ equals unity for a fully chaotic Gaussian source. The impurity in the sample used and finite momentum resolution can suppress the value of $\lambda$-parameter. In addition to this the non-Gaussian form of the correlation function and the FSI between particles can affect (suppress or enhance) its value. The last term in Eq.~(\ref{ll_eq}) is introduced to take into account the long tail observed in the measured data, where $a_{\rm res}$ is the residual amplitude and $r_{\rm res}$ is the width of the Gaussian.  

When the amplitude $a_{\rm res}$ in Eq.~(\ref{ll_eq}) is made to vanish, a fit performed on data causes a larger $\chi^2/{\rm NDF}$ (dashed line in Fig.~\ref{fig:CF_true}) and also the obtained $r_0$ is much smaller than the expected $r_0$ from previous measurements~\cite{pionHBT, kaonkaon, plambda}, which suggests that the measured correlation is wider than what the fit indicates in this scenario. This effect can be explained by the presence of a negative residual correlation in the data, which is expected to be wider than the correlation from the parent particles. Therefore, to include the effect of a residual correlation, a Gaussian term $a_{\rm res}\exp(-Q^2 r_{\rm res}^2)$ is incorporated in the correlation function (solid line in Fig.~\ref{fig:CF_true}). A negative residual correlation contribution is required with $a_{\rm res} = -0.044 \pm 0.004^{+0.048}_{-0.009}$ and $r_{\rm res} = 0.43 \pm 0.04^{+0.43}_{-0.03}$ fm, where the first error is statistical and the second is systematic.  Such a wide correlation could possibly arise from residual correlations caused by decaying parents such as $\Sigma^0$ and $\Xi$, and coupling of $N\Xi$ to the $\Lambda\Lambda$ channel. The fit parameters obtained with the residual correlation term are  $N = 1.006 \pm 0.001$, $\lambda = 0.18 \pm 0.05^{+0.12}_{-0.06}$, $a_0 = -1.10 \pm 0.37^{+0.68}_{-0.08}$ fm, $r_{\rm eff} = 8.52 \pm 2.56^{+2.09}_{-0.74}$ fm and  $r_0 = 2.96 \pm 0.38^{+0.96}_{-0.02}$ fm with $\chi^2/{\rm NDF}$ = 0.56. All the systematic errors on the parameters are uncorrelated errors. The Gaussian term is empirical and its origin is not fully understood. However, the addition of this term improves fit results and the obtained $r_0$ is compatible with expectations. The LL analytical model fit to data suggests that a repulsive interaction exists between $\Lambda\Lambda$ pairs, whereas the fit to the same data from K. Morita {\it et al.} showed that the  $\Lambda\Lambda$ interaction potential is weakly attractive~\cite{Ohnishi}. The conclusion about an attractive or a repulsive potential is limited by our statistics and is model dependent. However, all model fits to data suggest that a rather weak interaction is present between $\Lambda\Lambda$ pairs. 

\begin{figure}[hbt]
\begin{center}

\epsfxsize = 3.4in
\epsfysize = 2.7in
\epsffile{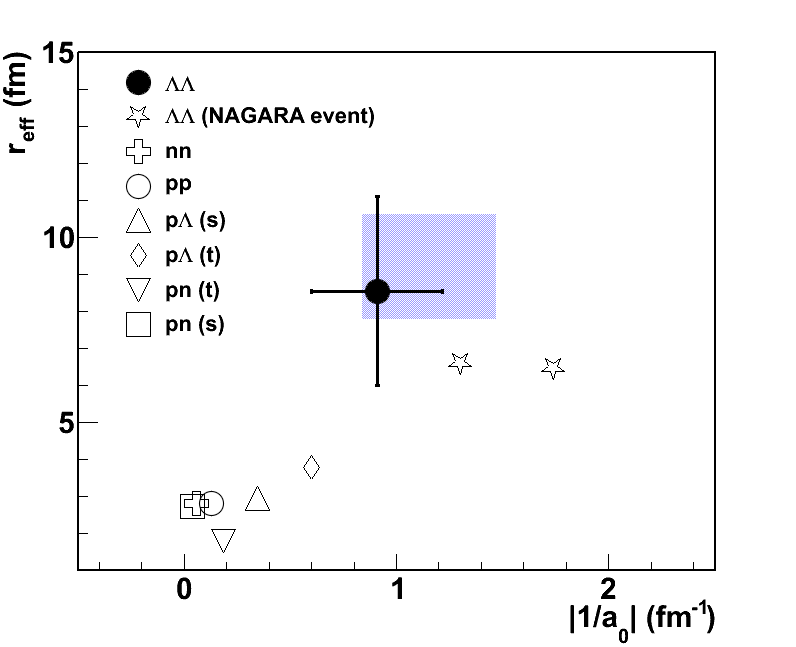} 
\end{center}
\caption{ 
(color online). The $\Lambda\Lambda$ interaction parameters from this experiment (solid circle), where the shaded band represents the systematic error. The interaction parameters from $pp$, $pn$ singlet ($s$) and triplet ($t$) states, and from $nn$, $p\Lambda$ ($s$) and $p\Lambda$ ($t$) states are shown as open markers~\cite{IP}. Also, the $\Lambda\Lambda$ interaction parameters which reproduce the NAGARA event are shown as open stars~\cite{Nagara_Hiyama, Nagara_FG}.
}
\label{fig:IP_lm}
\end{figure}

The scattering length and the effective radius obtained from the model fit are shown in Fig.~\ref{fig:IP_lm}. For comparison, interaction parameters for $pp$, $nn$ and $pn$ singlet ($s$) and triplet ($t$) states as well as for $p\Lambda$ singlet ($s$) and triplet ($t$) states are also shown in Fig.~\ref{fig:IP_lm}~\cite{IP}. It is observed that $|a_{\Lambda\Lambda}| < |a_{p\Lambda}| < |a_{NN}|$. The LL analytical model gives a negative $a_0$ parameter and favors a slightly repulsive interaction in our convention which is different from a weak attractive potential extracted from the NAGARA event and the KEK result~\cite{KEK, Nagara_Hiyama, Nagara_FG}. The fit parameters are still limited by statistics and our fitted $a_0$ is 1.6$\sigma$ from a sign change. A negative sign for the scattering length (in our convention) is a necessary though not sufficient condition for the existence of a $\Lambda\Lambda$ bound state.  

If a $\Lambda\Lambda$ resonance exists near the threshold, that would induce large correlations between two $\Lambda$s at small relative momentum~\cite{Muller,scott}. For the $\Lambda\Lambda$ system below the $N\Xi$ and $\Sigma\Sigma$ thresholds ($k < 161$ MeV/$c$), the FSI effect is included in the correlation function through the s-wave amplitude~\cite{sakurai}, 

\begin{equation}
\label{fsi}
f(k)=\frac{1}{k\cot\delta - ik}\,, 
\end{equation}
\noindent
where $k$ and $\delta$ are relative momentum and $s$-wave phase shift, respectively. The effective-range approximation for $k\cot\delta$ is

\begin{equation}
\label{kcot}
k\cot\delta = \frac{1}{a_0}+r_{\rm eff} \frac{k^2}{2}.
\end{equation}
\noindent

Equation~(\ref{fsi}) should satisfy the single-channel unitarity condition, ${\rm Im} f(k) = k|f(k)|^2$, with real parameters $a_0$ and $r_{\rm eff}$. When the scattering amplitude is saturated by a resonance, it can be re-written~\cite{landau} in the form   

\begin{equation}
\label{ad}
f(k) = \frac{1}{(k_0^2 - k^2)/(2\mu\gamma) - ik}
\end{equation}

\noindent
Comparing the above to Eqs.~(\ref{fsi}) and (\ref{kcot}), one sees that 
$1/a_{0} = k_{0}^{2}/(2\mu\gamma)$ and $r_{\rm eff} = -1/\mu\gamma$, where $k_0$, $\mu$ and $\gamma$ are the relative momentum where the resonance occurs, the reduced mass, and a positive constant, respectively. The scattering length (effective range) becomes positive (negative) so that the $k\cot\delta$ term vanishes at $k=k_0$~\cite{Hyodo}. The signs of $a_0$ and $r_{\rm eff}$ obtained from the fit to our data contradict Eq.~(\ref{ad}), which suggests the non-existence of a $\Lambda\Lambda$ resonance saturating the $s$-wave below the $N\Xi$ and $\Sigma\Sigma$ thresholds. More discussion on the  existence of $H$ as a resonance pole can be found in~\cite{Ohnishi}.   

Assuming that $H$-dibaryons are stable against strong decay of $\Lambda$, and are produced through coalescence of $\Lambda\Lambda$ pairs, the yield for the $H$-dibaryon can be related to the $\Lambda$ yield by 
$d^2 N_H / 2\pi p_T dp_T dy = 16B ( d^2 N_{\Lambda} / 2\pi p_T dp_T dy )^2$, 
where $B$ is a constant known as the coalescence coefficient. From pure phase space considerations, the coalescence rate is proportional to $Q^3$~\cite{cole}. For a weakly bound or deuteron-like bound state $H$, the $\Lambda\Lambda$ correlation below the coalescence length $Q$ would be depleted.  Our data show no depletion in the correlation strength in our measured region, which indicates that the value of $Q$ at coalescence for the $H$ dibaryon, if it exists, must be below 0.07 GeV/$c$ where we no longer have significant statistics.  Therefore, because the deuteron coalescence coefficient $B =(4.0 \pm 2.0)\times 10^{-4}$ (GeV/$c)^2$~\cite{dueteron_CR,lambda_yield} for a $Q$ of approximately 0.22 GeV/$c$, we estimate that the $H$ dibaryon must have $B$ less than $(1.29 \pm 0.64) \times 10^{-5}$ (GeV/$c)^2$ for $Q < 0.07$ GeV/$c$. The corresponding upper limit for $p_T$-integrated $dN_H/dy$ is $(1.23 \pm 0.47_{\it{stat}} \pm 0.61_{\it{sys}})\times 10^{-4}$ if the coalescence mechanism applies to both the deuteron and the hypothetical $H$ particle.

In summary, we report the first measurement of the $\Lambda\Lambda$ correlation function in heavy-ion collisions, for Au+Au at $\sqrt{s_{NN}}= 200$ GeV.  The  measured correlation strength at $Q = 0$, $C(Q=0)$ is greater than 0.5 (the expectation from quantum statistics alone). In addition to the normal $\Lambda\Lambda$ correlation function, a Gaussian term is required to fit the data, possibly due to residual correlations.  The extracted Gaussian source radius is compatible with the expectation from previous measurements of pion, kaon and p$\Lambda$ correlations~\cite{pionHBT, kaonkaon, plambda}. The model fits to data suggest that the strength of the $\Lambda\Lambda$ interaction is weak. Numerical analysis of the final-state interaction effect using an $s$-wave scattering amplitude suggests the non-existence of a $\Lambda\Lambda$ resonance saturating the $s$-wave below the $N\Xi$ and  $\Sigma\Sigma$ thresholds. A limit on the yield of a deuteron-like bound $H$-dibaryon is also reported. 

\section{Acknowledgments}
We thank S. Pratt for helpful discussions.  We thank the RHIC Operations Group and RCF at BNL, the NERSC Center at LBNL, the KISTI Center in Korea, and the Open Science Grid consortium for providing resources and support. This work was supported in part by the Offices of NP and HEP within the U.S. DOE Office of Science, the U.S. NSF, CNRS/IN2P3, FAPESP CNPq of Brazil,  the Ministry of Education and Science of the Russian Federation, NNSFC, CAS, MoST and MoE of China, the Korean Research Foundation, GA and MSMT of the Czech Republic, FIAS of Germany, DAE, DST, and CSIR of India, the National Science Centre of Poland, National Research Foundation (NRF-2012004024), the Ministry of Science, Education and Sports of the Republic of Croatia, and RosAtom of Russia.

\renewcommand{\bibfont}{\small}

\end{document}